# Chemical Bonding Analysis on Amphoteric Hydrogen – Alkaline Earth Ammine Borohydrides

S Kiruthika[1, 2] and P Ravindran[1, 2, 3, 4, a)]

[1]*Department of Physics, Central University of Tamil Nadu, Thiruvarur-610005*
[2]*Simulation Center for Atomic and Nanoscale Materials, Central University of Tamil Nadu, Thiruvarur-610005*
[3]*Department of Materials Science, Central University of Tamil Nadu, Thiruvarur-610005*
[4]Department of Chemistry, Center for Materials Science and Nanotechnology, University of Oslo, P.O. Box 1033 Blindern, N 1035 Oslo, Norway

[a)] email: raviphy@cutn.ac.in

**Abstract.** Usually the ions in solid are in the positive oxidation states or in the negative oxidation state depending upon the chemical environment. It is highly unusual for an ion having both positive as well as negative oxidation state in a particular compound. Structural analysis suggest that the alkaline earth ammine borohydrides (AABH) with the chemical formula *M* $(BH_4)_2(NH_3)_2$ (*M* = Mg, Ca, or Sr) where hydrogen is present in +1 and -1 oxidation states. In order to understand the oxidation states of hydrogen and also the character of chemical bond present in AABH we have made charge density, electron localization function, Born effective charge, Bader effective charge, and density of states analyses using result from the density functional calculations. Our detailed analyses show that hydrogen is in amphoteric behavior with hydrogen closer to boron is in negative oxidation state and that closer to nitrogen is in the positive oxidation state. Due to the presence of finite covalent bonding between the consitutents in AABH the oxidation state of hydrogen is non-interger value. The confirmation of the presence of amphtoric behavior of hydrogen in AABH has implication in hydrogen storage applications.

## INTRODUCTION

Hydrogen is one of the ideal and promising environmental benign energy carriers. However developing safe and efficient hydrogen storage materials for mobile applications is a key challenge. Though hydrogen can be stored in gas, liquid and solid forms the solid states storage based on light weight element is the only solution to meet the long term goal with respect to the gravimetric hydrogen density. Recently identified chemical hydrides are the potential energy storage materials due to their favorable gravimetric and volumetric density[1,2]. But the key challenge is the slow kinetics and poor reversibilty added with high decomposition temperture. Alkaline earth amine metal borohydrides (AAMBs) possess the properties of both metal borohydrides as well as ammonia borane. Hence, these compounds exhibit a gripping properties like high hydrogen capacity and favorable dehydrogenation properties.

In most of the hydrogen storage materials hydrogen is in negative oxidation state and less than 10% of the known hydrides were hydrogen is in positive oxidation states. Further if the hydrogen is in negative or positive oxidation state in solids, due to Coulomb repulsion of same charged ions the hydrogen atoms are present with the interatomic distance 2 Å or more. This limit the volumetric density of hydrogen in solids. If one can identify that hydrogen in both negative as well as positive oxidation states then due to Coulomb attraction of hydrogen ions one can increase the gravimetric as well as volumetric density of hydrogen. This will pave the way to identify potential hydrogen storage materials. So, we investigated AABH where experimental structural analysis suggested that hydrogen is in amphoteric state using state of the art computational methods.

## COMPUTATIONAL DETAILS

The total energy calculation has been performed according to the projector augmented wave method (PAW) as implemented in the Vienna ab initio simulations package (VASP). The calculation was made by utilizing density functional theory, employing the generalized gradient–approximation (GGA) functional of Perdew et.al[3]. All the calculation were carried out with a 300 eV plane wave cutoff. The structural optimization were performed using force and stress minimization. For the structural optimization the k-points were generated using the Monkhorst pack method with the grid of $5\times 6\times 2$ for Mg $(BH_4)_2(NH_3)_2$ and similar k-point density where used for other compounds considering the present study.

## RESULTS AND DISSCUSSION

The crystal structure of $M(BH_4)_2(NH_3)_2$ (M=Mg, Ca, Sr) are illustrated in figure (1); all three compound has orthorhombic structure with space group Pcab, Pbcn and Pnc2 for $Mg(BH_4)_2(NH_3)_2$, $Ca(BH_4)_2(NH_3)_2$, and $Sr(BH_4)_2(NH_3)_2$, respectively[4,5]. In all the three compounds the alkaline earth metal is coordinated octahedrally by four bridging $(BH_4)^-$ group in the plane and two $NH_3$ groups bonded to axially over the plane.

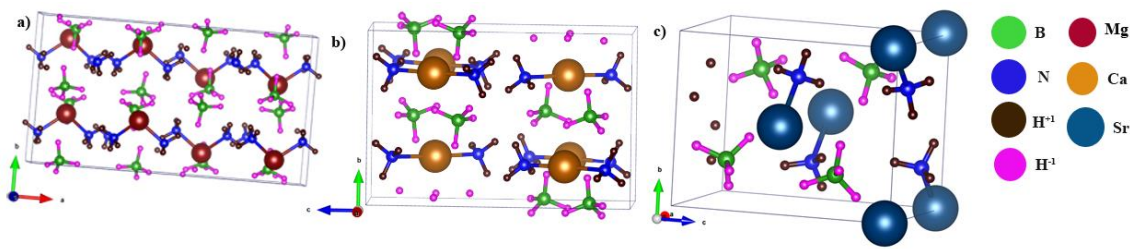

**FIGURE 1.** Crystal structure of 1(a) $Mg(BH_4)_2(NH_3)_2$, 1(b) $Ca(BH_4)_2(NH_3)_2$, and 1(c) $Sr(BH_4)_2(NH_3)_2$.

In order to understand the chemical bonding present between the constituents in M $(BH_4)_2(NH_3)_2$ density of states (DOS) and charge density distribution were investigated. The total DOS and partial DOS (PDOS) for AABH are shown in the figure (2). From figure (2) it is clear that all these compounds are insulators since there is large band gap present (5.3 eV, 5.3 eV, and 5.4 eV for M = Mg, Ca and Sr respectively) between the valance band maximum and conduction band minimum.

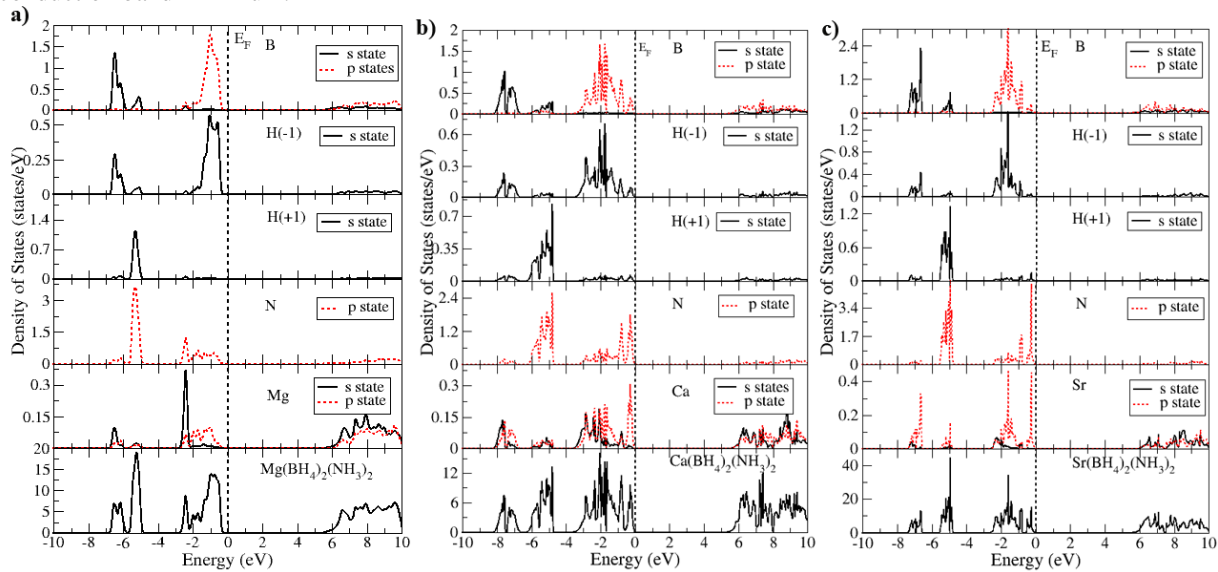

**FIGURE 2.** Total DOS (last row) and partial DOS (first five rows) of M $(BH_4)_2(NH_3)_2$ where M=Mg, Ca, Sr are given in a, b, and c respectively.

The PDOS of all three compounds show similar features: the lowest energy region is dominated by B *2s* and $H^{-1}$ *1s* states; the next band is primarily composed of N *2p* and $H^{+1}$ *1s* hybridized states; the top the valance band is composed of B *2p* and $H^{-1}$ *1s* hybridized states. From the PDOS in figure (2) it is clearly evident that hydrogen is in two different oxidation state in consistent with the experimental suggestion based on structural analysis. The energetically degenerate nature of B 2p and $H^{-1}$ 1s states as well as N 2p and $H^{+1}$ 1s states indicate the presence of noticeable covalent bonding $B-H^{-1}$ and $N-H^{+1}$.

In order to substantiate the amphoteric behavior of hydrogen and also the presence finite covalent bonding between hydrogen with neighbors in AABH we have shown charge density distribution for $Mg(BH_4)_2(NH_3)_2$ in Fig.(3) as representative system. Similar kind of charge density distribution is seen for both $Ca(BH_4)_2(NH_3)_2$ and $Sr(BH_4)_2(NH_3)_2$. The negligibly small charge density distribution around Mg indicating that Mg is in 2+ oxidation state. There is finite charge distributed between hydrogen with neighbors and also the directional distribution of charge density indicating the presence of finite covalent bonding. However the hydrogen closer to the B has relatively more spherical distributed charge with small charge between B and $H^{-1}$ indicating noticeable ionic bonding character. So, one can conclude that the bonding interaction between $N-H^{+1}$ and $B-H^{-1}$ is iono-covalent nature. Due to nearly protonic state of hydrogen near N the interatomic distance is short and the charge is mostly accumulated in the N site with in the $NH_3$ structural sub unit. In contrast the hydrogen neighboring B is in -1 oxidation state and hence the interatomic distance is large and substantial amount of charge is present in the H site.

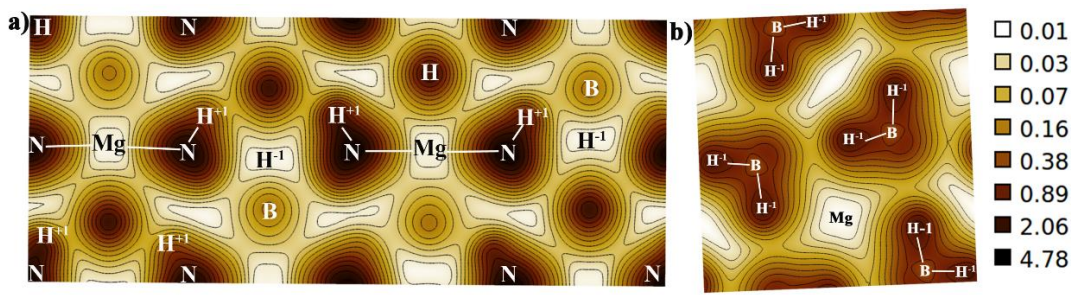

**FIGURE 3.** Charge density distribution for $Mg(BH_4)_2(NH_3)_2$ where plane representing N-H bond and B-H bond are shown in (a) and (b) respectively.

For further understanding of chemical bonding between constituents and the oxidation state of hydrogen we have calculated the Born effective charge (BEC) and their values are tabulated in Table (1). The BEC of B has a smaller value than the corresponding nominal ionic charge (+3). Similarly, the effective charge value of H-1 is smaller than its nominal ionic charge (-1). The smaller value of BEC indicating that this systems is not having pure ionic bonding. The presence of finite value in the off-diagonal components of BEC tensor indicating covalency. Further the diagonal element of the BEC also not the same for all three direction suggesting the presence of anisotropic charge distributions and hence covalent bonding. Interestingly as expected from DOS, charge density and structural analysis the hydrogen closer to B has negative BEC value and that closer to N has the positive BEC value confirming the presence of amphoteric hydrogen in AABH.

**TABLE 1.** The Born Effective charge

| Compound Name | Ion | $Z^*$ | | | | | | | | |
|---|---|---|---|---|---|---|---|---|---|---|
| | | xx | yy | zz | xy | yz | zx | xz | zy | yx |
| $Mg(BH_4)_2(NH_3)_2$ | Mg | 1.783 | 2.296 | 2.302 | 0.006 | 0.162 | -0.004 | -0.013 | -0.158 | -0.006 |
| | B1 | 0.131 | -0.036 | -0.032 | 0.010 | -0.122 | 0.016 | 0.001 | 0.122 | -0.026 |
| | $H^{-1}$ | -0.219 | -0.389 | -0.208 | -0.012 | -0.105 | -0.004 | -0.002 | -0.077 | -0.013 |
| | N1 | -1.115 | -0.943 | -0.966 | 0.025 | -0.014 | -0.059 | -0.005 | 0.011 | 0.031 |
| | $H^{+1}$ | 0.468 | 0.320 | 0.322 | -0.064 | 0.084 | -0.004 | -0.061 | 0.064 | -0.061 |
| $Ca(BH_4)_2(NH_3)_2$ | Ca | 2.286 | 2.358 | 2.074 | -0.000 | 0.000 | 0.266 | 0.247 | -0.000 | -0.000 |
| | B1 | -0.034 | 0.041 | 0.015 | 0.094 | 0.028 | -0.136 | -0.115 | 0.067 | 0.070 |
| | $H^{-1}$ | -0.216 | -0.240 | -0.170 | -0.082 | 0.009 | 0.031 | 0.049 | -0.016 | -0.098 |
| | N1 | -1.005 | -0.817 | -1.090 | 0.07 | -0.002 | 0.072 | 0.045 | -0.001 | 0.113 |
| | $H^{+1}$ | 0.289 | 0.365 | 0.379 | -0.010 | 0.086 | 0.031 | -0.061 | 0.125 | 0.060 |
| $Sr(BH_4)_2(NH_3)_2$ | Sr | 2.153 | 2.332 | 2.406 | 0.268 | 0.000 | 0.000 | 0.000 | 0.000 | 0.250 |
| | B1 | 0.051 | 0.023 | 0.093 | -0.107 | -0.080 | -0.030 | -0.001 | -0.044 | -0.072 |
| | $H^{-1}$ | 0.440 | -0.216 | 0.337 | -0.044 | -0.071 | 0.121 | 0.124 | -0.070 | -0.060 |
| | N1 | -1.018 | -0.934 | -0.704 | 0.091 | -0.040 | 0.020 | 0.028 | -0.086 | 0.088 |
| | $H^{+1}$ | 0.317 | 0.419 | 0.243 | 0.048 | 0.106 | 0.111 | 0.058 | 0.130 | 0.155 |

We have calculated the Bader's effective charge using the atom in molecule concept by topological analysis of charge density. The Bader charge analysis reveals the strong ionic character of the Mg/Ca/Sr $(NH_3BH_4)_2$ bonds since the metal ion is seen to have donated most of its electron toward the $(NH_3)_2$ & $(BH4)_2$ groups. The metal (M=Mg, Ca, Sr) atoms possess essentially the same charge, with the slight difference of 0.14, 0.06 e being probably related to the differences in bond lengths or more generally in their respective crystal structure. The calculated Bader effective charge given in table (2) shows that the alkaline earth metal sites have the effective charges smaller than the corresponding ionic charge (i.e.2+). Concerning the hydrogens atoms in Mg $(BH_4)_2(NH_3)_2$, one can see two different hydrogen atom one bonded to N that donated a significant part of their electrons (donated +0.55 e); and the hydrogen atoms, bonded to the boron accept significant amount of electrons (accepted -0.55e). This once again confirming the amphoteric behavior of hydrogen in AABH.

**TABLE 2.** The Bader effective charge

| Compound Name | Atom | Bader Effective Charge |
|---|---|---|
| $Mg(BH_4)_2(NH_3)_2$ | Mg | 1.6206 |
| | N | -1.6254 |
| | $H^{+1}$ | 0.5584 |
| | $H^{-1}$ | -0.5444 |
| | B | 1.4043 |
| $Ca(BH)_4(NH_3)_2$ | Ca | 1.4842 |
| | N | -1.5339 |
| | $H^{+1}$ | 0.5036 |
| | $H^{-1}$ | -0.5595 |
| | B | 1.5120 |
| $Sr(BH_4)_2(NH_3)_2$ | Sr | 1.5417 |
| | N | -1.5219 |
| | $H^{+1}$ | 0.5287 |
| | $H^{-1}$ | -0.5692 |
| | B | 1.5585 |

## CONCLUSION

Using Ab- initio band structure method we have analyzed the chemical bonding between the constituents in *M* $(BH_4)_2(NH_3)_2$ (*M* = Mg, Ca, or Sr). We have found that the alkaline earth metals are mainly in purely ionic states where as the bonding between N-H and B-H is of iono-covalent nature. All these compounds are found to be insulator. The Hydrogen in these compounds are exits in amphoteric condition. However due to covalency effect the hydrogen site charge is not integer value and this is much smaller than pure ionic case (i.e. +0.55 or -0.55). As we have demonstrated here it is possible to have hydrogen storage system where one can keep positively and negatively charged hydrogen with in the same structural framework. If one can keep positive and negatively charged hydrogen ions adjacent to each other they can be placed much closer due to Coulomb attraction and hence improve the hydrogen storage capacity. WE hope that the present study motivated further research in this direction by identifying practical hydrogen storage materials those possess amphoteric hydrogen.

## ACKNOWLEDGEMENT

The authors are grateful to the Department of Science and Technology, India for the funding support via Grant No. SR/NM/NS-1123/2013 and the Research Council of Norway for computing time on the Norwegian supercomputer facilities.

## REFERENCES

1. [1] H. Wu, W. Zhou, and T. Yildirim, J. Am. Chem. Soc. **130**, 14834 (2008).
2. [2] X. Chen and X. Yu, J. Phys. Chem. C **116**, 11900 (2012).
3. [3] J.P. Perdew, K. Burke, and M. Ernzerhof, Phys. Rev. Lett. **77**, 3865 (1996).
4. [4] L.H. Jepsen, Y.S. Lee, R. Radovan Cerny, R.S. Sarusie, Y.W. Cho, F. Besenbacher, and T.R. Jensen, ChemSusChem **8**, 3472 (2015).
5. [5] G. Soloveichik, J. Her, P.W. Stephens, Y. Gao, J. Rijssenbeek, M. Andrus, J. Zhao, G. Soloveichik, J. Her, P.W. Stephens, Y. Gao, and J. Rijssenbeek, ChemSusChem **8**, 3472 (2015).